\documentclass{aa}

\usepackage{graphicx}
\usepackage{tablefootnote}
\usepackage{longtable}
\usepackage{lscape}

\usepackage{txfonts}
\usepackage{supertabular,booktabs}

\usepackage{hyperref}
\hypersetup{
    colorlinks = true,
    citecolor ={blue}
}

\usepackage{tabularx} 
\newcommand\clearrow{\global\let\rowmac\relax} 
\clearrow
\usepackage[normalem]{ulem}
\usepackage{soul,xcolor}    

\begin{document}

   \title{A comprehensive search for hot subdwarf stars using Gaia and TESS
   }
   \subtitle{l. Pulsating hot subdwarf B stars}

   \author{Murat Uzundag\inst{1},
          Jurek Krzesinski\inst{2},
          Ingrid Pelisoli\inst{3}, 
          P\'eter N\'emeth\inst{4,5},
          Roberto Silvotti\inst{6},
          Maja Vu\v{c}kovi\'{c}\inst{7}\\
          Harry Dawson\inst{8},
          Stephan Geier\inst{8}
          }

   \institute{Institute of Astronomy, KU Leuven,
    Celestijnenlaan 200D, 3001, Leuven, Belgium
   \\\email{\url{muratuzundag.astro@gmail.com}}
         \and
    Astronomical Observatory, Jagiellonian University, ul. Orla 171, PL-30-244 Krakow, Poland
    \and
    Department of Physics, University of Warwick, Gibbet Hill Road, Coventry CV4 7AL, UK
    \and
    Astronomical Institute of the Czech Academy of Sciences, CZ-251 65, Ond\v{r}ejov, Czech Republic
    \and 
    Astroserver.org, F\H{o} t\'{e}r 1, 8533 Malomsok, Hungary
    \and
    INAF-Osservatorio Astrofisico di Torino, strada dell'Osservatorio 20, 10025 Pino Torinese, Italy
    \and
    Instituto de Física y Astronomía, Universidad de Valparaíso, Gran Bretaña 1111, Playa Ancha, Valparaíso 2360102, Chile
    \and 
    Institute for Physics and Astronomy, University of Potsdam, Karl-Liebknecht-Str. 24/25, 14476 Potsdam, Germany 
    }
\date{}

    \titlerunning{A comprehensive search for pulsating hot subdwarf stars}
    \authorrunning{Murat Uzundag et al.}

 
  \abstract
   {Hot subdwarf B (sdB) stars are evolved, subluminous, helium-burning stars, most likely formed when red-giant stars lose their hydrogen envelope via interactions with close companions. They play an important role in our understanding of binary evolution, stellar atmospheres, and interiors.
   Within the sdB population, only a small fraction are known to exhibit pulsations. 
   Pulsating sdBs have typically been discovered serendipitously in various photometric surveys, lacking specific selection criteria for the sample. Consequently, while individual properties of these stars are well-known, a comprehensive understanding of the entire population and many related questions remain unanswered. The introduction of Gaia has presented an exceptional chance to create an unbiased sample by employing precise criteria and ensuring a high degree of completeness. 
   The progression of high-precision and high-duty cycle photometric monitoring facilitated by space missions such as Kepler/K2 and the Transiting Exoplanet Survey Satellite (TESS) has yielded an unparalleled wealth of data for pulsating sdBs.
   In this work, we created a dataset of confirmed pulsating sdB stars by combining information from various ground- and space-based photometric surveys.
   Utilizing this dataset, we present a thorough approach to search for pulsating sdB stars based on the current Gaia DR3 sample.  
   Using TESS photometry, we discovered 61 new pulsating sdB stars and 20 variable sdBs whose source of variability remains to be determined through future spectroscopic follow-up observations.
   
   }


   \maketitle
%

\section{Introduction}
Hot subdwarf B-type stars, referred to as sdB stars, are thought to be core-helium (He) burning stars characterized by an extremely thin hydrogen (H) envelope, with an envelope mass less than 0.02 times the mass of the Sun, $M_{\rm env}$ $<$ 0.02 $M_{\odot}$ \citep[e.g.,][]{2022ApJ...933..137G}. 
Their average mass is in close proximity to the mass at which core-He flash occurs, approximately around 0.47 times the mass of the Sun \citep{2012A&A...539A..12F}.
These sdB stars represent evolved, compact objects with surface gravities ($\log{g}$) in the range of 5.2 to 6.2 dex and effective temperatures ($T_{\rm eff}$) ranging from 20\,000 to 40\,000 K \citep{1994ApJ...432..351S, 2008ASPC..392...75G}.
 They reside in the region between the main sequence and the cooling stage of white dwarfs, referred to as the extreme horizontal branch (EHB) stars \citep[see][for a detailed overview]{2009ARA&A..47..211H, heber2016}.
The evolutionary history of sdB stars involves significant mass loss, predominantly induced by binary interactions during the late stages of the red giant phase \citep{han2002, han2003}. This phase results in the loss of almost all of the H-rich envelope, leaving behind a core that burns He but possesses an envelope too thin to sustain H-shell burning. Throughout a duration of approximately $10^{8}$ years, sdB stars continue to burn He in their cores. Following the depletion of He in their core, they transition into a phase where He is burned in a shell surrounding a core composed of carbon and oxygen (C/O), transforming into subdwarf O (sdO) stars. Ultimately, these stars conclude their lifecycle as white dwarfs \citep[e.g.,][]{dorman1993}.

\citet{kilkenny1997} made the pioneering discovery of rapid pulsations in a subset of hot sdB stars, which have come to be known as V361 Hya stars and are often referred to as short-period sdB variable stars. The V361 Hya stars exhibit multiperiodic pulsations characterized by periods ranging from 60\,s to 800\,s. Within this frequency range, these pulsations correspond to low-degree, low-order pressure(p)-modes. These modes are associated with photometric amplitudes that can reach up to several percent of the stars' mean brightness, as corroborated by studies conducted by \citet{reed2007}, \citet{ostensen2010}, and \citet{green2011}.
The excitation of these modes is attributed to a classical $\kappa$-mechanism, primarily driven by the accumulation of iron group elements, with iron itself being a dominant contributor, in a region known as the $Z$-bump. This mechanism was elucidated by \citet{charpinet1996} and further elaborated upon by \citet{charpinet1997}. These studies demonstrated that radiative levitation plays a crucial role as a physical process that enhances the concentrations of iron group elements. This enhancement is a prerequisite for the activation of the pulsational modes.
The p-mode sdB pulsators are typically located within a temperature range spanning from 28\,000 K to 35\,000 K, with surface gravities characterized by values in the range of $\log g$ 5 to 6 dex. Following this discovery, a different class of sdB pulsators known as V1093 Her stars, exhibiting long-period pulsations, was identified by \citet{green2003}. These stars display variations in brightness with periods that can extend to a few hours and have amplitudes on the order of 0.1 percent or less of their mean brightness \citep[e.g.,][]{reed2011}.
The oscillation frequencies observed in these pulsators are linked to g-modes characterized by low-degree ($l <$ 3) and medium- to high-order (with 10 $<$ $n$ $<$ 60), 
a phenomenon driven by the same $\kappa$-mechanism, caused by the accumulation of iron group elements, as outlined by \citet{fontaine2003} and further discussed by \citet{charpinet2011}. 
In contrast to the p-mode sdB pulsators, the g-mode sdB pulsators have somewhat cooler temperatures, ranging from 22\,000 K to 30\,000 K, and their surface gravities ($\log{g}$) typically fall within the range of 5.0 dex to 5.5 dex.
Within the spectrum of pulsating sdB stars, there exists a subset referred to as "hybrid" sdB pulsators, which exhibit both g- and p-modes concurrently \citep{schuh2006}. These hybrid sdB pulsators occupy a distinct region in the Hertzsprung-Russell (HR) diagram, positioned between the regions associated with p-mode and g-mode sdB pulsators, as depicted in figure 5 of \citet{green2011}. These objects hold particular significance as they provide a unique opportunity to conduct asteroseismic investigations, allowing for the study of both the core structure and the outer layers of sdBVs.

Both the Kepler mission \citep{borucki2010} and the TESS mission \citep{Ricker2014} have significantly contributed to identifying new candidates of pulsating sdBs. During the primary Kepler mission, a group of 18 pulsating sdB stars were observed. 
Among these stars, the majority, specifically 16 of them, were identified as long-period g-mode pulsators, while the remaining two exhibited short-period p-mode pulsations \citep{ostensen2010,ostensen2011,kawaler2010,baran2011,reed2011}.
Additionally, within the old open cluster NGC6791, three known sdB stars were also discovered to exhibit pulsations \citep{pablo2011, reed2012}.
During the secondary Kepler mission known as K2 mission \citep{haas2014}, about 50 sdBs were identified as pulsators, and ongoing analyses continue to study these stars. To date, around 20 of these sdBVs have been published, along with detailed information about their atmospheric parameters \citep{reedk22016, kernk22017, bachulskik22016, reedk22019, barank22019, silvottik22019, reedk22020, ostensen2020, Ma2022, Ma2023}.
Currently, the analysis of data from the TESS mission is ongoing, and a detailed analysis of 13 long-period sdBVs were reported  \citep{charpinet2019, reed2020, saho2020, Uzundag2021, Silvotti2022, Uzundag2023}.

\citet{Holdsworth2017} compiled a table containing information on 110 known pulsating sdBs up to that point. In a subsequent review, \citet{Lynas2021} provided a thorough overview, including 56 documented pulsating sdB stars.
More recently, \citet{Grootel2021}, \citet{KB2022} and \citet{Baran2023} have made notable contributions by confirming and discovering new pulsating sdB stars including short- and long-period ones through observations conducted via the TESS mission. 

In this study, an extensive dataset was meticulously assembled, encompassing all documented pulsating sdBV stars. As of now, this comprehensive compilation incorporates records for 256 pulsating sdBV stars. 
Utilizing this existing dataset, we developed a robust selection algorithm outlined in Sect. \ref{sample_selection}, which is based on the Gaia color-magnitude diagram.
In Sect. \ref{observations}, we gathered and analyzed all accessible light curves from TESS, including both short and ultra-short-cadence observations. Then, we generated periodograms for over 2,300 objects.
In Section \ref{Var_search}, our focus shifted towards the classification of newly identified members.
In Sect. \ref{spec}, we presented the spectroscopic observations and produced the atmospheric parameters of newly discovered sdBs and a comparison with the established pulsating sdB population.
Finally, we summarized our results and conclusions in Sect. \ref{conclusion}.

\section{Sample selection}
\label{sample_selection}

Pulsating sdBs have mostly been found by chance in photometric campaigns, with no clearly defined selection criteria for the sample. This means that, while individual parameters can be very well determined, an understanding of the population as a whole is missing and many open questions remain such as the pulsation occurrence rate.
The advent of Gaia \citep{2018A&A...616A...1G} provided an unparalleled opportunity to assemble an unbiased sample by enabling the selection of candidates through well-defined criteria and a high level of completeness. \citet{Culpan2022} successfully compiled a catalog of 61\,585 potential hot subdwarf candidates utilizing Gaia EDR3 \citep{2021A&A...649A...1G}.

In the pursuit of identifying potential pulsating sdB candidates, an initial step involved the comprehensive compilation of all known stars from existing literature. This inclusive subset constituted 256 pulsating sdBs, visually represented by red dots in Fig. \ref{fig:HR_sky}. We provided a catalog comprising all known pulsating sdBs in Appendix Table \ref{known_tab}. Subsequently, a distribution plot was constructed based on the known pulsating sdB stars as shown in the right panel in Fig. \ref{fig:HR_sky}.  
The mean value of the distribution is indicated by the horizontal dashed black line at $M_{G} =  4.18$. 
Then, in order to encompass all known pulsating stars, we considered the +/-$\sigma$ range around the mean of the distribution. Therefore, the upper limit at $+2\sigma$ is $M_{G} = 6.11$, marked by the upper horizontal black line, while the lower limit at $M_{G} = 2.26$, represented by the lower horizontal black line.
To refine the selection, composite sdB systems such as sdBVs+F/G/K were excluded, specifically those positioned beyond $G_{BP}-G_{RP} > 0.1$ as denoted by the vertical black line (the future work will focus on the search for pulsating sdBs in binary systems). 
Following these refined criteria, an initial identification yielded a dataset of 25\,615 stars that is shown in Fig. \ref{fig:HR_sky} with a blue shaded area. 

In Fig. \ref{fig:HR_sky}, we show the color-magnitude diagram of selected pulsating sdB candidates within the blue shaded area (left panel) along with known pulsating sdBs and their spatial distribution with respect to the galactic coordinate system (right panel).

\begin{figure*}
    \includegraphics[clip,width=1.0\columnwidth]{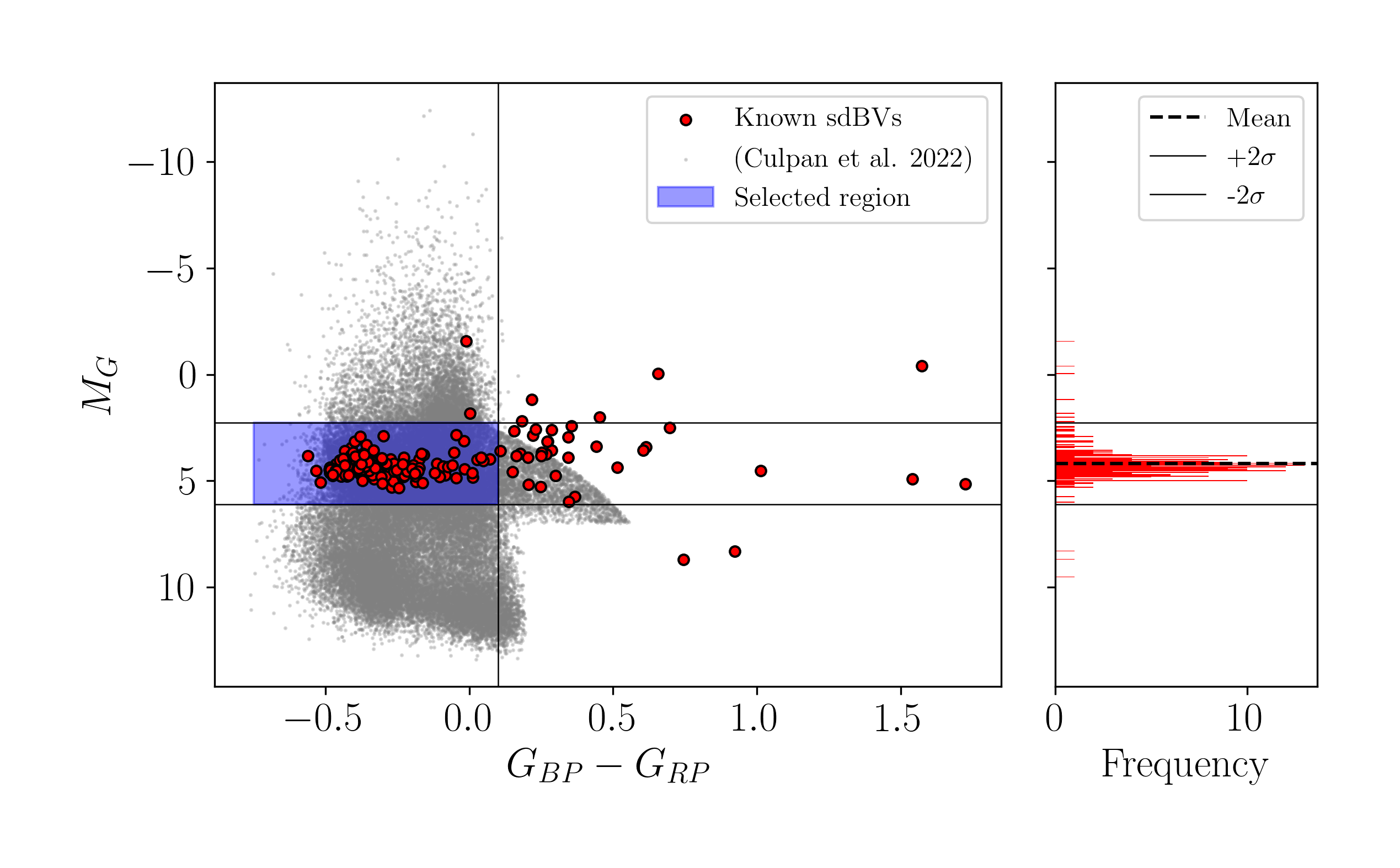} 
    \includegraphics[clip,width=1.0\columnwidth]{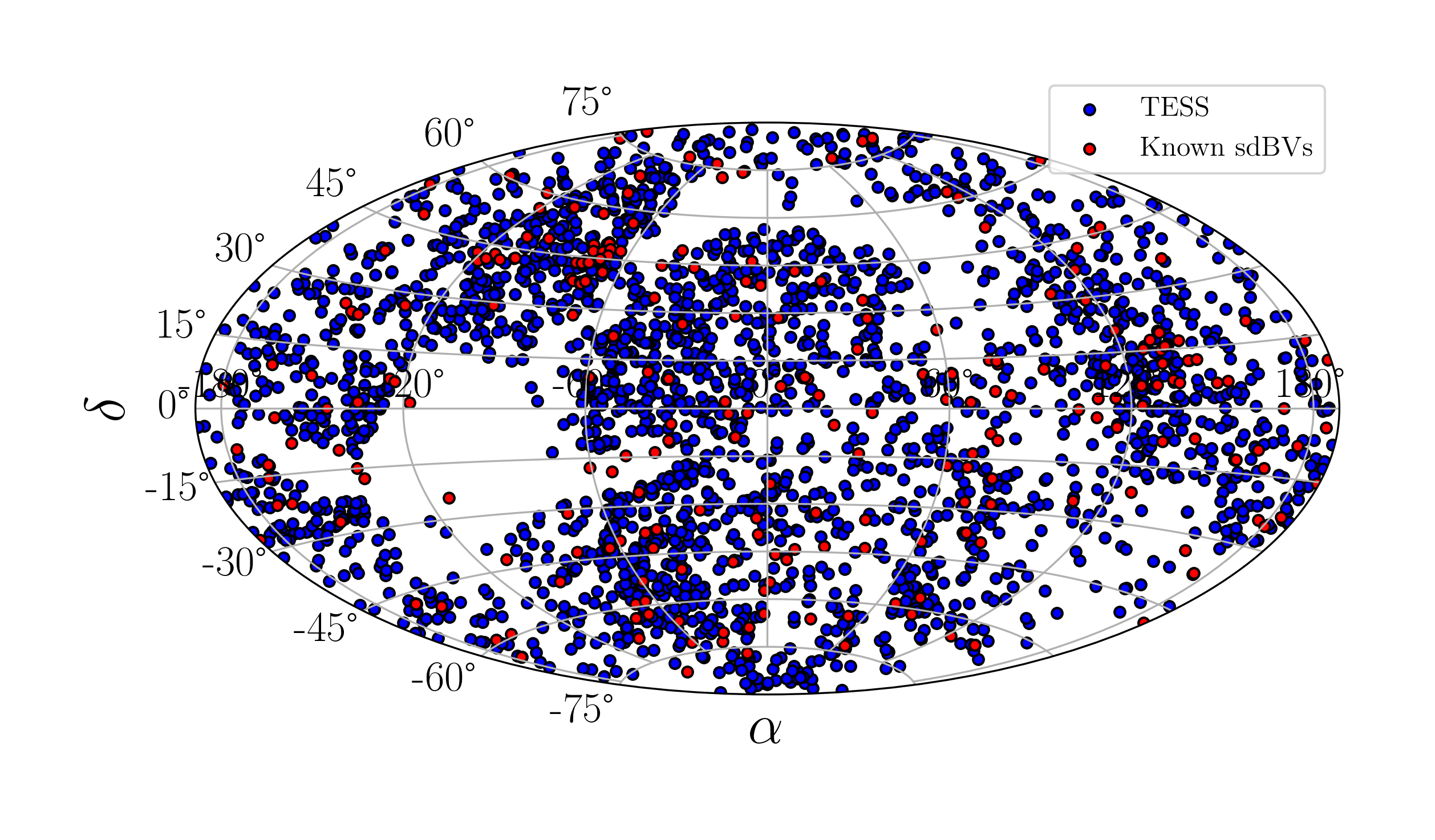} 
 \caption{{\sc Left:} 
 The color-magnitude diagram of identified potential hot subdwarf candidates with grey dots from \cite{Culpan2022}. On top of that we highlighted 256 known sdBVs in red, compiled from the literature.  
 A distribution plot (middle panel) is constructed based on the known pulsating sdBs.  The mean value of the distribution is indicated by the horizontal dashed black line while the upper and lower limits are shown by horizontal black lines.
 Our refined sample which excludes composite binaries is visualized as a blue shaded box in the color-magnitude diagram.
         {\sc Right:} Sky locations (Galactic coordinates, Aitoff projection) of the selected pulsating sdB candidates that are observed with TESS and known pulsating sdBs with respect to the galactic coordinate system using the same color coding.  }  
    \label{fig:HR_sky}
\end{figure*}

By limiting the sample to $G_{mag} < 17$ mag considering TESS' detection limit, we identified 10\,452 targets. 
However, we noted that setting a upper limit of $G_{mag} < 17$ could result in the exclusion of a few pulsators exhibiting lower amplitudes.
Lastly, to obtain all observed stars with TESS, we crossmatched the catalog with the TESS input catalog \citep{Stassun2019}, and the final sample was restricted to 2\,371 objects with TESS light curves. 

\section{TESS observations}
\label{observations}

TESS is actively observing numerous sdBs across the entire sky in consecutive sectors. Each sector is observed for a duration of approximately 27 days. As of now, data from all the sectors (labeled from 1 to 68) covered by the TESS mission are accessible. 

TESS data is organized and stored in the Mikulski Archive for Space Telescopes (MAST)\footnote{\url{http://archive.stsci.edu}} as target pixel files (TPFs) in FITS format. The light curves are accessible in two processed forms: calibrated using Simple Aperture Photometry (SAP), and pre-conditioned through Pre-search Data Conditioning Simple Aperture Photometry (PDCSAP). For this work, we make use of PDCSAP, which was processed using the Science Processing Operations Center (SPOC) pipeline \citep{Jenkins2016}. The pipeline is based on the Kepler Mission science pipeline and made available by the NASA Ames SPOC center and at the MAST archive. 

The TESS light curves were obtained from MAST in two cadences: 2-minute (short cadence) light curves, available for all 2\,371 hot subdwarf candidates, and 20-second (ultra-short cadence) light curves, applicable to 808 out of the 2\,371 targets. Given the extensive scope of this search for stellar variability, no specific effort was invested in optimizing pipeline apertures.

Additionally, TPFs of interest were retrieved from the MAST archive, managed by the Lightkurve Collaboration \citep{lightkurve2020}, for all candidates exhibiting variability. These TPFs comprise an 11x11 grid of pixels extracted from one of the four CCDs per camera where the target is located.

The TPFs were used to analyze fluxes in individual pixels of the pipeline apertures, when the source of variability was uncertain. To gauge this uncertainty, we utilized the crowding factor, represented by the keyword $\tt{CROWDSAP}$. This factor estimates the contamination level of the PDCSAP flux, providing the ratio of the target to the background star flux in the pipeline aperture, accounting for the presence of potentially bright sources near the target.

\section{Variability search and classification method}
\label{Var_search}

In the following steps, we computed Lomb-Scargle periodograms \citep{2018ApJS..236...16V} for each target light curve. The code to perform this task was based on the python lightkurve package. To identify variability, a periodogram detection threshold of 5 sigma was applied. The frequency range for the detection threshold calculation extended from 0 to 360 d$^{-1}$ for a 2-minute cadence and 0 to 2\,160 d$^{-1}$ for a 20-second cadence. 

Variability in an object was evaluated by analyzing the maximum periodogram signal. If this signal surpassed the detection threshold, we proceeded to examine its frequency. In the initial step, we eliminated all stars displaying long-term variability and signals below the detection threshold, resulting in a subset of 1\,517 objects for subsequent analysis. 

It is important to note that sdBV pulsations are not expected at frequencies below 5 d$^{-1}$; therefore, any objects displaying maximum signals below 5\,d$^{-1}$ were excluded from the sample. Additionally, objects exhibiting a strong signal (exceeding the periodogram detection threshold by a factor of 5) within the frequency range of 5 to 35 d$^{-1}$ were systematically removed. This step aimed to eliminate potential binary systems from the sample. However, it is worth noting that compact binaries like ZTF J213056.71+442046.5, with an orbital frequency of approximately 37 d$^{-1}$ \citep{2020ApJ...891...45K}, could still be included in the sample of single pulsators. In fact, we identified two such objects: TIC107548305 and TIC~367090060, both exhibiting single peaks at frequencies of 40.833767 d$^{-1}$ and 37.725309 d$^{-1}$. In both cases however, the presence of sub-harmonic frequencies in their periodograms, as well as the analysis of the phase-folded light curves, led us to conclude that the real period of these binaries is twice as long and corresponds to the sub-harmonic frequencies. Therefore both stars were removed from our sample. While this procedure may inadvertently exclude some binaries featuring a pulsating primary sdB component, the primary focus remained on identifying single pulsators. 

At this stage, our sample was refined to 284 objects whose periodograms met the previously mentioned criteria and underwent visual inspection. This inspection resulted in the removal of several clearly identified binary or eclipsing binary light curves, as well as objects with periodograms indicating sub- or harmonics to the main peaks. Consequently, we arrived at a final sample comprising 61 objects classified as sdBV stars, as detailed in Table \ref{tab1}, and 20 sdBs of uncertain type documented in Table \ref{tab2}. The latter category encompasses stars displaying single peaks in their light curve periodograms and an absence of sub- and harmonics. Some periodograms of these stars exhibit potential signals below the 5-sigma but above the 4-sigma detection threshold. Due to the inherent difficulty in classification, most of these objects were labeled as binaries (B) or potential pulsating sdBs (G, I, P, H - depending on the pulsation type, as outlined below).

\begin{figure}
   \includegraphics[clip,width=1.0\columnwidth]{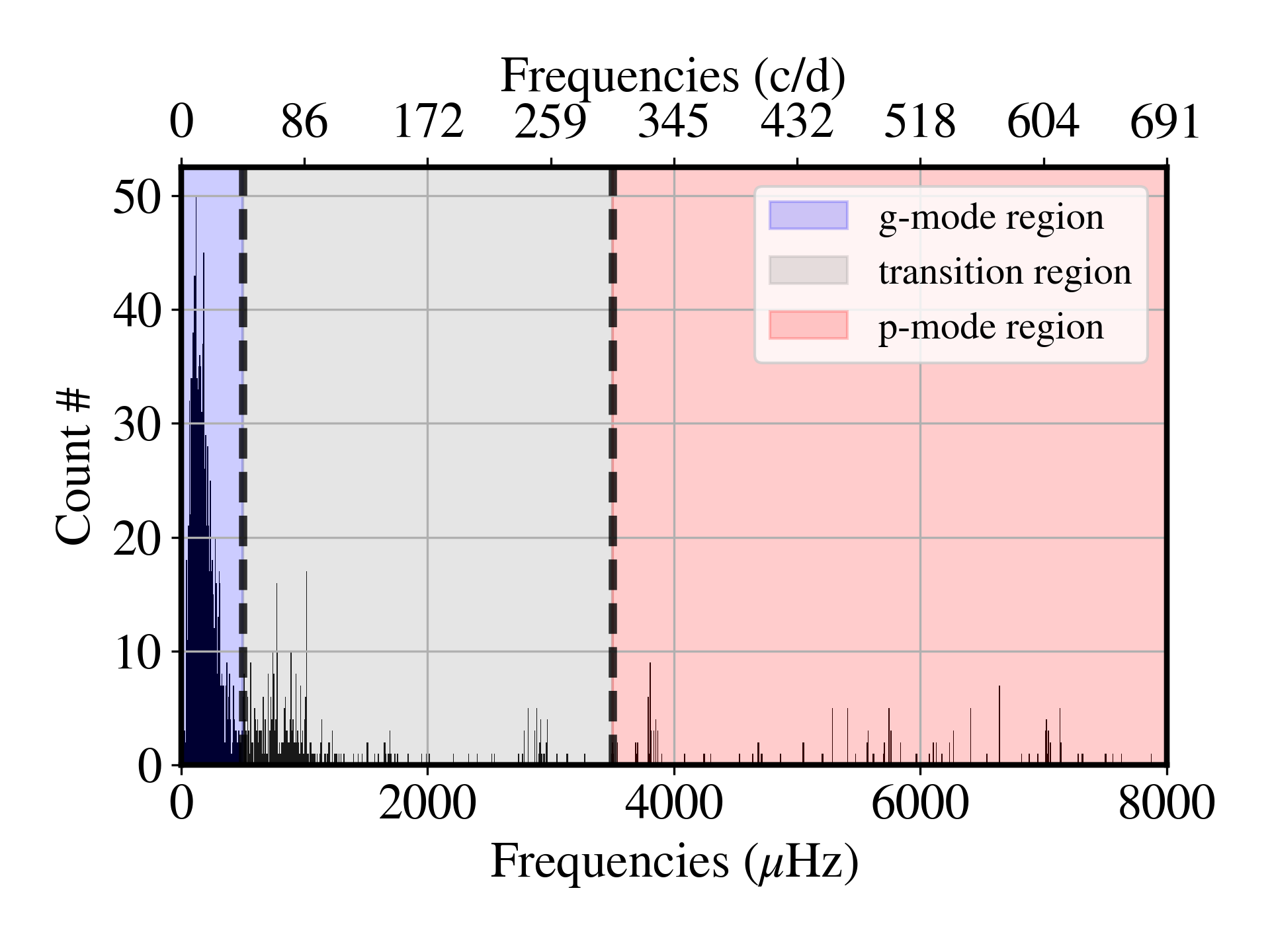} 
 \caption{Frequency distribution of pulsating sdBs which were observed by Kepler/K2. Theoretical boundaries for g- and p-mode regions are represented by the vertical black dashed lines. See text for more details. }
    \label{fig:F_dist}
\end{figure}

The classification of pulsation types (indicated in the "Type" column of tables \ref{tab1} and \ref{tab2}) should be based on a theoretical division into g- and p-modes, determined by the pulsation periods of the stars. As can be observed, the theoretical works by \cite{charpinet2000, fontaine2003, 2006MNRAS.372L..48J, 2009A&A...508..869H, 2014A&A...569A.123B}, provided excitation region plots for g- and p-mode sdB pulsators in the period vs. $T_{\rm eff}$ space. 
  While the period boundaries for g- and p-modes in these sources varied, one can infer that the common range for g-mode frequencies extends approximately from 4 to 55 d$^{-1}$ (equivalent to periods 21\,600 - 1\,570 seconds, see Fig. \ref{fig:F_dist} blue shaded region), and the p-mode excitation region can be safely assumed above 340 d$^{-1}$ (periods shorter than 250 seconds, as illustrated in Fig. \ref{fig:F_dist} pink shaded region). For frequencies falling between these boundaries (Fig. \ref{fig:F_dist} gray shaded region), classifying a pulsation mode based solely on observed frequencies in the Lomb-Scargle periodograms from photometric data alone is challenging.

To visualize the problem 
of classification
relying on photometric measurements, we conducted an analysis to determine the positions of pulsation frequencies within the amplitude diagram. This involved collecting periodicities from stars observed during the Kepler and K2 missions, excluding sdBVs found in binary systems. The dataset comprised 18  rich\footnote{By "rich pulsating sdBs", we refer to sdB pulsators that have substantial number of pulsation frequencies, allowing us to make use of asteroseismic methods and modelling.} pulsating sdBs\footnote{KIC10001893, KIC10139564, KIC10670103, KIC2437937, KIC2569576, KIC2697388, KIC2991276, KIC3527751, KIC5807616, KIC8302197, EPIC203948264, EPIC211779126, EPIC212707862, EPIC215776487, EPIC217280630, EPIC218717602, EPIC220641886, and EPIC248411044}, totaling 1\,630 individual frequencies identified in previous works. In Fig. \ref{fig:F_dist}, we illustrated the frequency distribution of this dataset.  As depicted, frequencies between $\sim$4 to 55 d$^{-1}$, corresponding to g-modes, constitute the most prevalent group of modes excited in sdBV stars. However, g-mode frequencies in Kepler sdBVs were found up to 86 d$^{-1}$, while p-modes begin to show above 172 d$^{-1}$. Frequencies between these two boundaries are considered intermediate modes.


 
As demonstrated, the theoretical division based on the common frequency ranges in all the mentioned theoretical works does not align precisely with the observed frequencies (e.g., the group of modes between 55-86 d$^{-1}$, which might be a continuation of g-modes, or the group of modes above 172 d$^{-1}$, which might belong to p-modes). This suggests that relying solely on photometry, one cannot provide accurate classification when a mode falls within the theoretical transition region.

To avoid confusion, in this paper, we choose to adhere to the classification based on the regions where theoretical works converge, deferring precise mode identification
to future investigations.
 Therefore, in Tables \ref{tab1} and \ref{tab2}, we classified stars as g-mode pulsators when the observed frequencies were between 5-55 d$^{-1}$ (labeled 'G' in the "Type" column), as p-modes ('P') when a periodogram frequency was above 340 d$^{-1}$, and as intermediate mode pulsators ('I') when the observed frequencies fell between 55-340 d$^{-1}$. To classify a star as a hybrid mode pulsator ('H'), it would need to meet both the G and P classification criteria. Furthermore, in the 8$^{th}$ column (Frequencies [d$^{-1}$], peaks - see Table \ref{tab1}), we provided the observed signal frequency ranges (in d$^{-1}$) and the corresponding number of signal peaks. Similarly, in Table \ref{tab2}, the 8$^{th}$ column includes frequencies of single peaks, along with details on sub- or harmonics and any additional frequencies present in the periodograms.

Periodograms for all new variable sdBs from both Tables \ref{tab1} and \ref{tab2} are presented in Fig. \ref{fig:Periodograms_a} for the 61 sdBVs and in Fig. \ref{fig:Periodograms_b} for the 20 sdBs of uncertain variability. Notably, Tables \ref{tab1} and \ref{tab2} reveals that some objects exhibit a low $\tt{CROWDSAP}$ factor suggesting possible flux contamination from background stars. Consequently, all objects with a $\tt{CROWDSAP}$ factor below 0.6 underwent additional scrutiny employing TPFs to assess target variability. Note that we did not attempt to recover the actual pulsation amplitudes, which are diminished due to the light pollution from the nearby stars affecting the targets. 


\section{Spectroscopic observations}
\label{spec}

The follow-up spectroscopic observations of the sdB pulsators analyzed in this study were conducted using three distinct instruments. These instruments include the Boller and Chivens (B\&C) spectrograph installed on the 2.5-meter (100-inch) Ir\'en\'ee du Pont Telescope\footnote{For a description of instrumentation, see: \url{http://www.lco.cl/?epkb_post_type_1=boller-and-chivens-specs}} at Las Campanas Observatory in Chile, the European Southern Observatory (ESO) Faint Object Spectrograph and Camera (v.2) (EFOSC2) \citep{buzzoni1984} mounted at the Nasmyth B focus of the New Technology Telescope (NTT) at La Silla Observatory in Chile, and the Southern Astrophysical Research (SOAR) Telescope, a 4.1-meter aperture optical and near-infrared telescope \citep{clemens2004}, located at Cerro Pach\'on, Chile.

We acquired low-resolution spectra to determine the atmospheric parameters, including the effective temperature ($T_{\rm eff}$), surface gravity ($\log{g}$), and helium (He) abundance. 
Two sdB stars, TIC152373379 and  TIC394678374, were observed with the B\&C spectrograph, while three sdBs, TIC269766236, TIC332841294 and TIC340223812, were observed with the EFOSC2. Only one sdB star, TIC181914779 was observed with the Goodman spectrograph. 

The B\&C spectra were obtained using the 600 lines/mm grating, corresponding to a central wavelength of 5\,000 \AA, and covering a wavelength range from 3\,427 to 6\,573 \AA. A 1 arcsec slit was utilized, resulting in a resolution of 3.1 \AA. 
In the case of EFOSC2 setup, the 6.4 $\AA$ 
resolution was obtained with a setup of grism \#7 and a 1 arcsec slit. In the case of the Goodman spectrograph, the 400 l/mm grating with the blaze wavelength 5\,500 \AA (M1:
3\,000-7\,050 \AA) with a slit of 1 arcsec was used and this setup provides a resolution of about 5.6 \AA. 

The data from B\&C and EFOSC2 were reduced and analysed using \textit{PyRAF}\footnote{\url{http://www.stsci.edu/institute/software_hardware/pyraf}} \citep{pyraf2012} procedures in the following way: 
first, bias correction and flat-field correction have been applied. 
Then, the pixel-to-pixel sensitivity variations were removed by dividing each pixel with the response function. 
After this reduction was completed, we applied wavelength calibrations using the frames obtained with the internal He-Ar comparison lamp. 
In a last step, flux calibrations were applied using the standard stars. 
The data reduction for Goodman has been partially done by using the instrument pipeline\footnote{\url{https://github.com/soar-telescope/goodman_pipeline}} including overscan, trim, slit trim, bias and flat corrections. 
For cosmic rays identification and removal, we used an algorithm as described by \citet{wojtek2004}, which is embedded in the pipeline. 
The extraction and calibration of the spectra were carried out similarly as for B\&C and EFOSC2 using standard \textit{PyRAF} tasks.
The details of the spectroscopic observations are given in Table \ref{obs_spec} including, instrument, date, exposure time and resolution.

\begin{table}
\setlength{\tabcolsep}{2.2pt}
\renewcommand{\arraystretch}{1.1}
\centering
\caption{Observing log of the spectroscopic data obtained for six pulsating sdB stars studied in this work. 
}
\begin{tabular}{ccccc}
\hline \hline
TIC &  Spectrograph & Date  & $t_{exp} $ & Resolution  \\
&          &  (UT)    &    (s)    &    ($\Delta \lambda$ (\AA)) \\ 
\hline
         
152373379   & B\&C   & 2019-08-21 & 450  &   3.1     \\
394678374   & B\&C   & 2019-08-21 & 480  &   3.1     \\
269766236   & EFOSC2 & 2019-06-20 & 350  &   6.4     \\
332841294   & EFOSC2 & 2019-06-13 & 375 &    6.4     \\
340223812   & EFOSC2 & 2019-06-19 & 375 &    6.4     \\
181914779   & GOODMAN& 2020-02-29 & 240  &   5.6     \\

\hline 
\label{obs_spec}
\end{tabular}
\end{table}

\subsection{Spectroscopic parameters}

We employed a steepest descent procedure implemented in {\sc XTgrid} \citep{2012MNRAS.427.2180N}, specifically designed for automating the spectral analysis of early-type stars utilizing {\sc Tlusty/Synspec} \citep{hubeny2017} non-Local Thermodynamic Equilibrium (non-LTE) stellar atmosphere models. 
This methodology employs an iterative chi-square minimization approach to fit observed data. 
Commencing with an initial model, the process converges on the best fit through successive approximations along the chi-square gradient. 
To reduce systematic effects like blaze function correction, flexure, or flux inconsistencies due to vignetting or slit-loss, the models undergo comparison with observations via a piecewise normalization.

For parameter determination of hot stars, both the completeness of included opacity sources and departures from LTE are crucial for accuracy. 
We determined that {\sc Tlusty} models, characterized by H, He composition, yield reliable results within the constraints of spectral resolution, coverage, and signal-to-noise of the survey data. 

\begin{figure*}
   \includegraphics[width=1.0\textwidth]{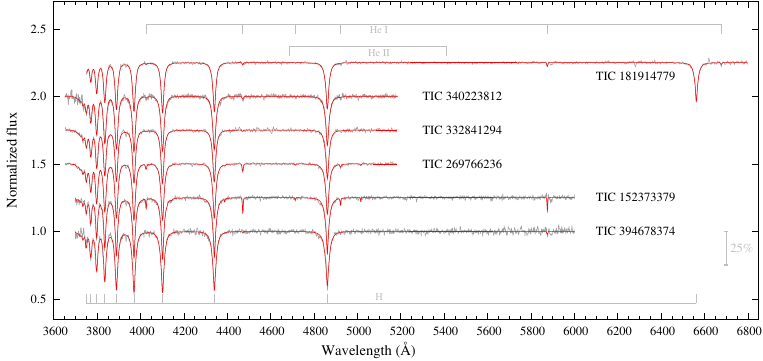} 
 \caption{Best-fit {\sc Tlusty/XTgrid} models (red) for the collection of continuum normalized spectra (grey) of the newly discovered variable sdB stars, organized by increasing $T_{\rm eff}$ from the bottom. 
 For a better representation the dentified
in the current investigation by spectroscopic observations,
while parameters for the remaining stars from Table 2 were
inferred from the literature.observed fluxes were adjusted to the final models and adjacent spectra are shifted by 25\%.} \label{fig:xtgrid}
\end{figure*}

{\sc XTgrid} dynamically computes the required {\sc Tlusty} atmosphere models and synthetic spectra, incorporating a recovery method to tolerate convergence failures and expedite convergence on a solution with a minimal number of models.
Figure\,\ref{fig:xtgrid} shows the best fits for the spectra of the six newly discovered variable sdB stars, and the surface parameters are listed at the top of Table\,\ref{Tab1:Sp_New}.  

Parameter errors are assessed by mapping the chi-square statistics around the solution. 
Parameters are altered in one dimension until the 60\% confidence limit is reached. 
Correlations near the best-fit values are incorporated into the final results for the surface temperature and gravity.

\begin{figure*}
   \includegraphics[width=1.0\textwidth]{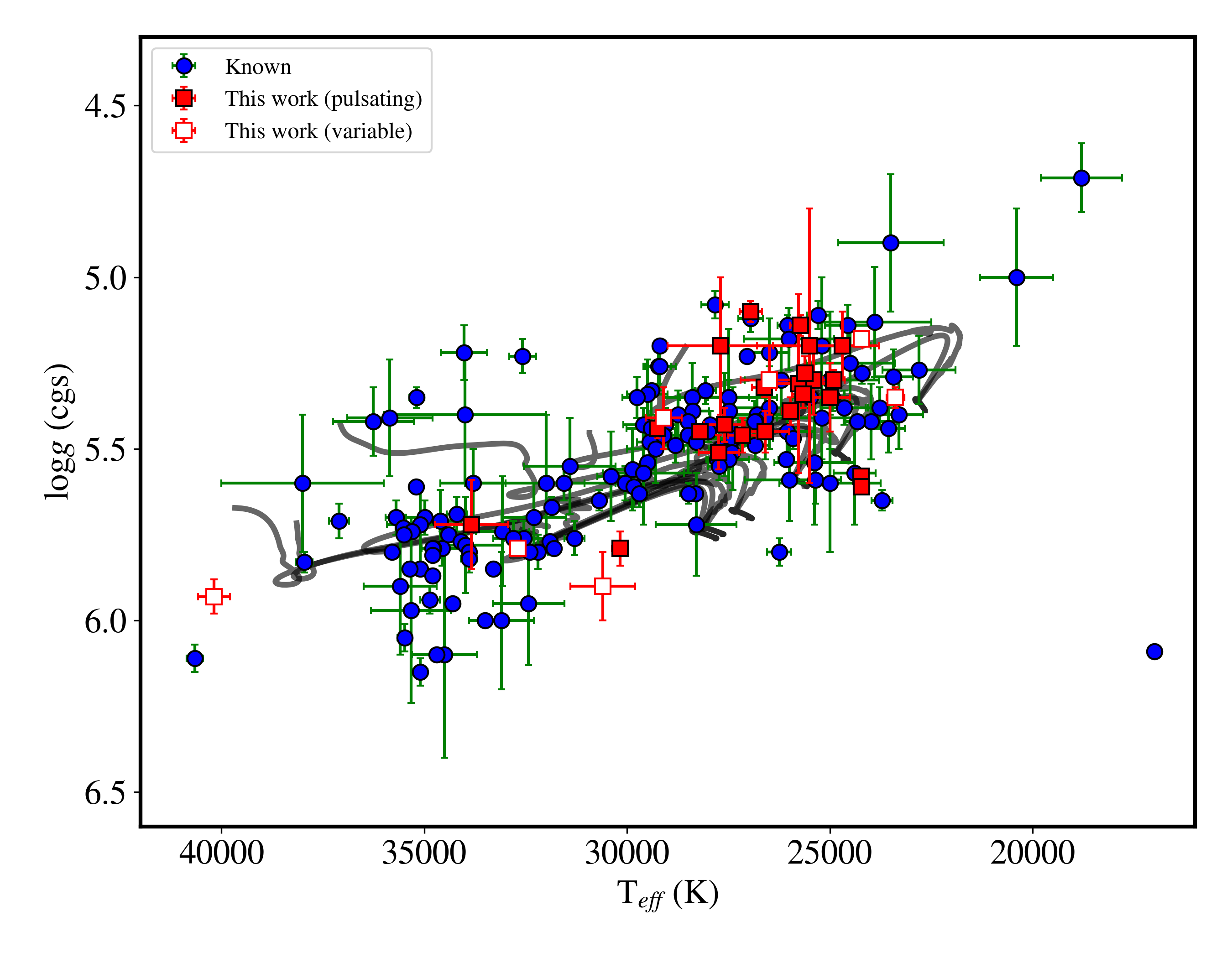} 
 \caption{ The position of the spectroscopically identified pulsating sdB star is indicated on the T$_{\rm eff} -$ log$g$ plane. Known pulsating sdBs are represented by blue dots, whereas newly discovered pulsating sdBs are denoted by red squares from Tab.\ref{tab1}. The open red squares present promising pulsating sdB candidates from Tab.\ref{tab2}. The evolutionary tracks have been adapted from \cite{Uzundag2021}. } \label{fig:Kiel}
\end{figure*}

\begin{figure*}
   \includegraphics[width=1.0\textwidth]{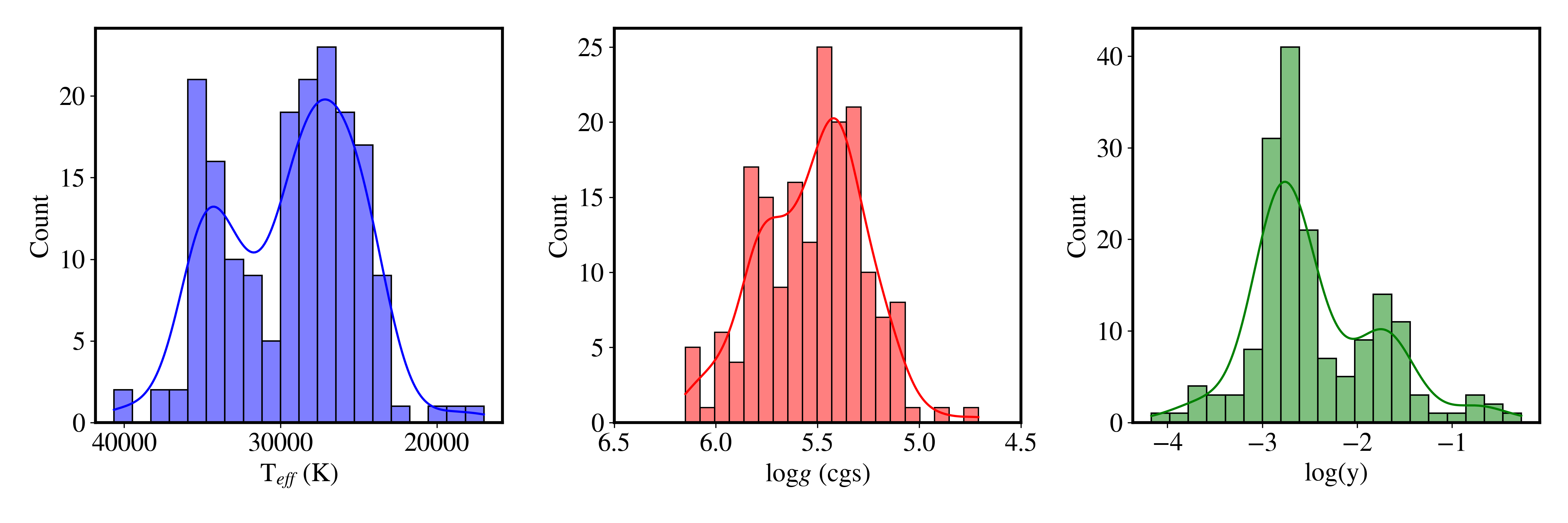} 
 \caption{The distribution plots display T$_{\rm eff}$ (first panel), log$g$ (second panel), and log(y) (third panel) for 179 pulsating sdBs for which atmospheric parameters are known.
 } \label{fig:hist_teff_loggy}
\end{figure*}

\begin{figure}
   \includegraphics[clip,width=1.0\columnwidth]{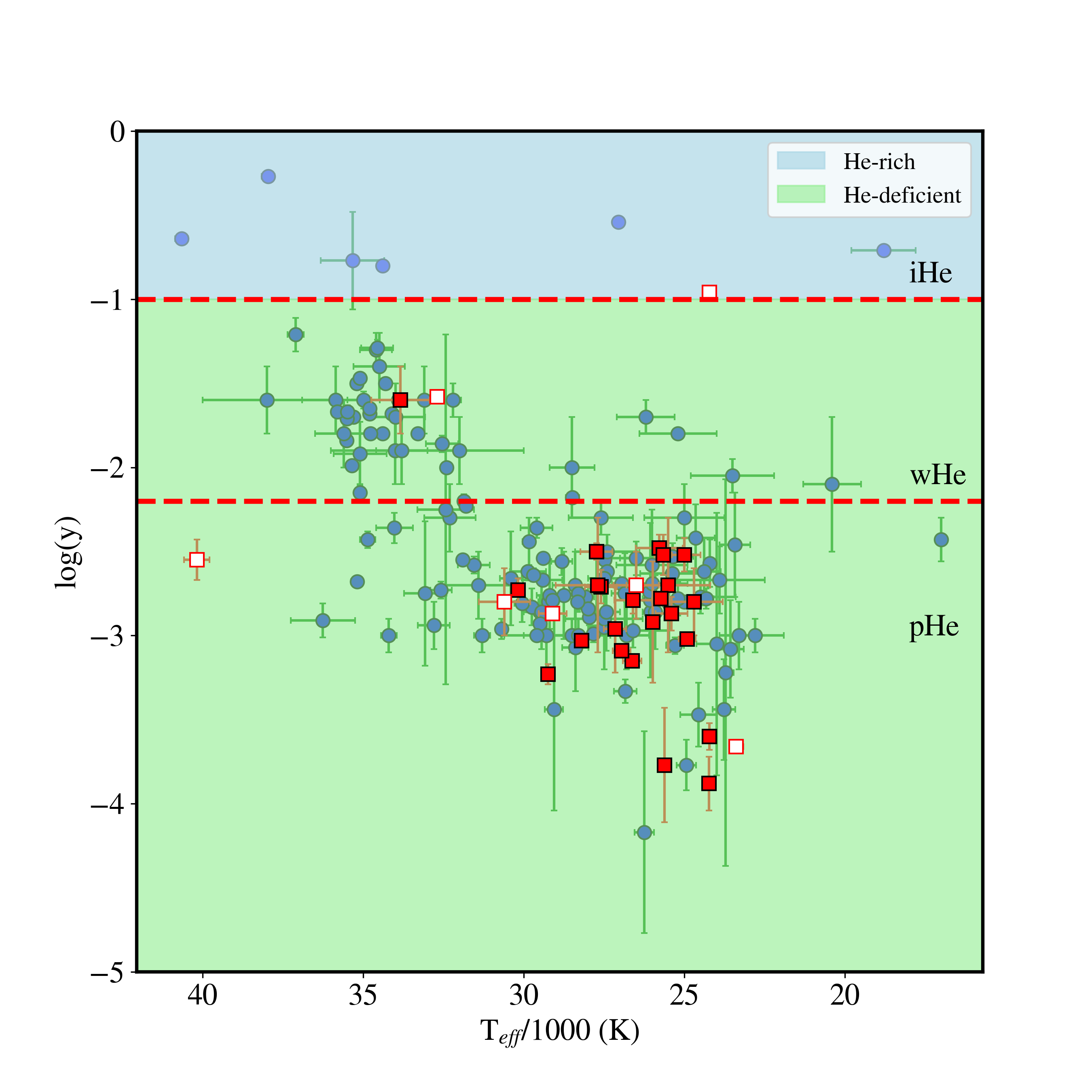} 
 \caption{T$_{\rm eff}$ versus log(y) diagram utilizes the classification scheme from \citet{2012MNRAS.427.2180N} to discern helium (He) abundance in pulsating sdB stars. The parameter space for sdBs is divided into He-rich (shaded light blue region) and He-deficient (shaded green area) stars. These divisions are subsequently categorized into He-poor (pHe), He-weak (wHe), and intermediate He-rich (iHe) classes. The color codes for the targets are the same as in Fig.\ref{fig:Kiel}.} \label{fig:teff_loggy}
\end{figure}


In this study, we have compiled a catalog comprising 256 known pulsating hot subdwarf stars that are listed in Table \ref{known_tab}. Within the existing literature, atmospheric parameters are available for 149 out of the total 256 stars included in our catalog. 
 In addition, in Table \ref{Tab1:Sp_New} we present the atmospheric parameters and Gaia distances for 22 of 61 newly discovered pulsating sdB stars presented in Table \ref{tab1}.  
The stars listed in Table \ref{Tab1:Sp_New} are arranged in order of their effective temperatures.
The initial five stars listed in the table are those identified in the current investigation by spectroscopic observations, while parameters for the remaining stars from Table \ref{Tab1:Sp_New} were inferred from the literature.  
The rest of the 39 stars that are listed in Table \ref{tab1} require additional spectroscopic observations. Our ongoing efforts are continuing, and in future work, we will present their atmospheric parameters.
Figure \ref{fig:Kiel} illustrates the newly found pulsating sdBs marked with red squares, while previously known ones are represented by blue dots. 

 Table \ref{Tab2:Sp_New} reports the atmospheric parameters for additional 9 of 20 new variable sdB stars listed in Table \ref{tab2} which variability we classified as uncertain. The first star TIC\,394678374 in Table \ref{Tab2:Sp_New} was identified in this work by spectroscopic observations.
Parameters for the remaining stars from Table \ref{Tab2:Sp_New} were
obtained from the literature. The location of most of them within the instability strip as depicted by the open red squares in Fig. \ref{fig:Kiel} suggests they are
potential pulsators.
However, the amplitude spectra in Fig. \ref{fig:Periodograms_c} show either a single peak or a few peaks, which could originate from binarity or pulsation. Notably, TIC\,125556577, positioned outside the g- and p-mode instability regions, may be a novel pulsating hot subdwarf O star. To confirm the nature of the variability, future photometric measurements for all targets in Table \ref{Tab2:Sp_New} are needed.

The evolutionary tracks from \cite{Uzundag2021}, based on stellar evolution models utilizing the $\tt LPCODE$ code \citep{althaus2005, millerbertolami2016}, are plotted in Fig. \ref{fig:Kiel}. These tracks, indicated by thick black lines, cover a mass range from $M_{\rm ZAHB}=0.46738$ to 0.473 $M_\odot$, with $\ell = 1$ g-mode frequencies computed using the adiabatic non-radial pulsation code $\tt LP-PUL$ \citep{Corsico2006a}.

In Fig. \ref{fig:hist_teff_loggy}, the distribution of all pulsating sdBs (179 stars) is depicted on the effective temperature, log$g$, and log(y) plane. As anticipated, the $T_{\rm eff}$ distribution (first panel) shows a bimodal pattern, segregating stars into g- and p-modes. The range of $T_{\rm eff}$ between 20\,000 K and 28\,000 K encompasses most g-mode dominated sdBs, while the higher $T_{\rm eff}$ range from 32\,000 K to 40\,000 K includes p-mode dominated stars. Hybrid stars are situated in these two regions, with hybridity observed across the entire span from 20\,000 K to 40\,000 K.
The log$g$ distribution exhibits a relatively narrow spread, with the majority of pulsating sdBs concentrated between 5 and 6. The distribution peaks at 5.55, indicating a central tendency within this log$g$ range.

In Fig.~\ref{fig:teff_loggy}, we used the T$_{\rm eff}$ versus log(y) diagram, employing the classification scheme from \citet{2012MNRAS.427.2180N} and later refined by \cite{2019ApJ...881....7L}, to distinguish He abundance in pulsating sdB stars. The sdB parameter space is split into He-rich (shaded light blue region) and He-deficient (shaded green area) stars. These divisions are then further classified into He-poor (pHe), He-weak (wHe), and intermediate He-rich (iHe) classes. The majority of pulsating sdB stars belong to the pHe class.
The two groups including g- and p-mode pulsating sdBs separate very well, in both T$_{\rm eff}$ vs. log$g$ and He abundance. 
P-mode pulsators are frequent among the hotter pHe stars, while g-mode is dominant among wHe stars. Moreover, the number of pulsating sdBs drops significantly in the He-rich range.

\begin{table*}
\begin{center}
\renewcommand{\arraystretch}{1.25}
\caption{Spectroscopic results of  22 newly discovered pulsating sdB stars from Tab. \ref{tab1} analyzed in this paper. }
\begin{tabular}{ccccccccc}
\hline \hline
TIC & $T_{\rm eff}$  & $\log{g}$  & $\log{(n_{\rm He} / n_{\rm H})}$ & Distance & Ref.  \\
    &        (K)           &      (cm\,s$^{-2}$)                    &                                  &    (pc)      &       \\    
\hline 
332841294   &   25620($\pm170$) & 5.28($\pm0.05$) & -3.77($\pm0.34$)  & $674 ^{+11}_{-13}$  &  (This work)     \\
269766236   &   25660($\pm270$) & 5.34($\pm0.04$) & -2.52($\pm0.12$)  & $637 ^{+16}_{-17}$  &  (This work)     \\
152373379   &   25780($\pm730$) & 5.31($\pm0.26$) & -2.48($\pm0.08$)  & $888 ^{+25}_{-22}$ &  (This work) \\
340223812   &   25980($\pm210$) & 5.39($\pm0.04$) & -2.92($\pm0.36$)  & $663 ^{+9}_{-11}$  &  (This work)     \\
181914779   &   29250($\pm100$) & 5.44($\pm0.03$) & -3.23($\pm0.06$)  & $518 ^{+12}_{-13}$  &  (This work)     \\
\hline
118297100   &   24230($\pm191$)	&	5.58($\pm0.02$) &	-3.88($\pm$0.16)	&  $523^{+13}_{-11}$   &    \citep{2021ApJS..256...28L}  \\
265787226   &   24700($\pm900$)	&	5.2($\pm0.1 $) &	-2.8 ($\pm$0.2) 	&  $920^{+25}_{-26}$   &    \citep{Edelmann2003}  \\
461658287   &   24911($\pm267$)	&	5.3($\pm0.03$) &	-3.02($\pm$0.04)	&  $2015^{+566}_{-350}$&    \citep{2021ApJS..256...28L}  \\
36707830    &   25000($\pm500$)	&	5.35($\pm0.1 $) &	-2.52($\pm$0.1) 	&  $461^{+12}_{-12}$   &    \citep{Kupfer2015}  \\
118032308   &   25400($\pm500$)	&	5.3	($\pm0.1 $) &	-2.87($\pm$0.1) 	&  $564^{+9}_{-10}$    &    \citep{Geier2010}  \\
259126140   &   25500($\pm1300$)&	5.2	($\pm0.4 $) &	-2.7 ($\pm$0.4) 	&  $1592^{+61}_{-71}$  &    \citep{Edelmann2003}  \\
292467033   &   25735($\pm251$)	&	5.14($\pm0.03$) &	-2.78($\pm$0.03)	&  $1072^{+49}_{-37}$  &    \citep{2021ApJS..256...28L}  \\
17561485    &   26596($\pm583$)	&	5.45($\pm0.06$) &	-2.79($\pm$0.08)	&  $1417^{+81}_{-118}$ &    \citep{2021ApJS..256...28L}  \\
138623536   &   26624($\pm293$)	&	5.32($\pm0.03$) &	-3.15($\pm$0.04)	&  $829^{+21}_{-19}$   &    \citep{2021ApJS..256...28L}  \\
429807453   &   26954($\pm274$)	&	5.1	($\pm0.03$) &	-3.09($\pm$0.03)	&  $654^{+12}_{-11}$   &    \citep{2021ApJS..256...28L}  \\
399171956   &   27160($\pm350$)	&	5.46($\pm0.05$) &	-2.96($\pm$0.26)	&  $440^{+18}_{-19}$   &    \citep{2012MNRAS.427.2180N}  \\
66493797    &   27600($\pm500$)	&	5.43($\pm0.05$) &	-2.71($\pm$0.1 )	&                      &    \citep{Geier2010}  \\
229593795   &   27700($\pm1300$)&	5.2	($\pm0.2 $) &	-2.7 ($\pm$0.4 )	&  $1597^{+78}_{-71}$  &    \citep{Edelmann2003}  \\
11489563    &   27738($\pm500$)	&	5.51($\pm0.05$) &	-2.5 ($\pm$0.05)	&                      &    \citep{Lisker2005}  \\
381203990   &   28202($\pm155$)	&	5.45($\pm0.02$) &	-3.03($\pm$0.02)	&  $554 ^{+10}_{-8}$   &    \citep{2021ApJS..256...28L}  \\
43965472    &   30176($\pm194$)	&	5.79($\pm0.05$) &	-2.73($\pm$0.07)	&  $474 ^{+8}_{-10}$   &    \citep{2021ApJS..256...28L}  \\
178081355   &   33841($\pm904$)	&	5.72($\pm0.13$) &	-1.6 ($\pm$0.2 )	&  $1104^{+71}_{-55}$  &    \citep{2017OAst...26..164G}  \\
\hline 
\label{Tab1:Sp_New}
\end{tabular}
\end{center}
\end{table*}

\begin{table*}
\begin{center}
\renewcommand{\arraystretch}{1.25}
\caption{Spectroscopic results of  9 newly discovered variable sdB stars analyzed in this paper from Tab.\ref{tab2}. }
\begin{tabular}{ccccccccc}
\hline \hline
TIC & $T_{\rm eff}$  & $\log{g}$  & $\log{(n_{\rm He} / n_{\rm H})}$ & Distance & Ref.  \\
    &        (K)           &      (cm\,s$^{-2}$)                    &                                  &    (pc)      &       \\    
\hline 
394678374   &   24220($\pm190$) & 5.61($\pm0.02$) & -3.60($\pm0.08$)  & $684 ^{+11}_{-12}$ &  (This work)     \\
345451496  &   23385($\pm217$)  &5.35($\pm0.03$)		&-3.66	($\pm0.04$)    &     $1271^{+79}_{-59}$  & \citep{2021ApJS..256...28L}    \\
165453703  &   24219($\pm72 $)  &5.18($\pm0.02$)		&-0.96	($\pm0.01$)    &     $689^{+16}_{-12}$  & \citep{2021ApJS..256...28L}    \\
242347840  &   26500($\pm700$)  &5.3 ($\pm0.1 $)		&-2.7	($\pm0.2 $)    &     $1051^{+48}_{-45}$  & \citep{Edelmann2003}    \\
286099192  &   29100($\pm440$)  &5.41($\pm0.09$)		&-2.87	($\pm0   $)    &     $1086^{+74}_{-57}$  & \citep{2012MNRAS.427.2180N}    \\
58873368   &   30600($\pm800$)  &5.9 ($\pm0.1 $)		&-2.8	($\pm0.2 $)    &     $1023^{+34}_{-36}$  & \citep{Edelmann2003}    \\
239122172  &   32693($\pm87 $)  &5.79($\pm0.02$)		&-1.58	($\pm0.01$)    &     $777^{+33}_{-29}$  & \citep{2021ApJS..256...28L}    \\
363766470  &   40179($\pm394$)  &5.93($\pm0.05$)		&-2.55	($\pm0.12$)    &     $923^{+20}_{-22}$  & \citep{2021ApJS..256...28L}    \\
125556577  &   55000($\pm3500$) &6   ($\pm0.4 $)		&-0.3	($\pm0.15$)    &     $1121^{+42}_{-41}$  &  \citep{Hunger1981}   \\
\hline 
\label{Tab2:Sp_New}
\end{tabular}
\end{center}
\end{table*}

\section{Summary, discussion and future prospects}
\label{conclusion}

In this study, we generated a dataset comprising confirmed pulsating sdB stars by consolidating data from different ground- and space-based photometric surveys. Employing this dataset, we conducted a comprehensive investigation to identify pulsating sdB stars within the existing Gaia DR3 sample. Through the analysis of TESS photometry for 2\,371 stars, we discovered 81 new variable sdB stars, comprising 61 pulsating ones (refer to Table \ref{tab1}) and 20 sdBs exhibiting uncertain variability (refer to Table \ref{tab2}). The specific nature of the variability in these 20 stars requires additional investigation, which will be carried out through subsequent spectroscopic and photometric follow-up observations.

We determined the atmospheric parameters for six stars by matching synthetic spectra to the recently acquired low-resolution spectra from Dupont/B\&C, NTT/EFOSC2, and SOAR/Goodman.
Additionally, we collected the atmospheric parameters from the literature for 17 stars.  
The newly discovered sdB pulsating stars exhibit effective temperatures ranging from 24\,220 K to 33\,841 K, and their surface gravity (log$g$) falls within the 5.1 to 5.8 dex range. 
This confirms that these stars distinctly belong to the g- and p-mode sdBV parameter space.

Concerning our mode classification presented in Tables \ref{tab1}, \ref{tab2}, the classification of pulsation modes based on Kepler sdBVs shown in Figure \ref{fig:F_dist}, would lead to the inference that gravity modes typically extend up to 86 d$^{-1}$, intermediate modes are situated between 86 - 172 d$^{-1}$, and acoustic p-modes are above 172 d$^{-1}$. Therefore, it would change our mode classification in both tables \ref{tab1} and \ref{tab2}. However, the classification derived from the frequency position in Figure \ref{fig:F_dist} is based on its occurrence according to a set of the best-investigated sdBV periodograms and constitutes rather an experimental or statistical approach, while more theoretical work in this regard is necessary.

Additionally, the objects presented in Table \ref{tab2} (Figure \ref{fig:Periodograms_c}) exhibit single frequencies (in a couple of cases, two frequencies) located in the low-frequency regions, i.e., below 40 d$^{-1}$. This frequency region also corresponds to the binary orbital frequency range; however, in the absence of indications of sub- or harmonic frequencies and phased light curves not showing any clear evidence of binarity, differentiation between binary and pulsating g-modes is impossible based solely on photometric observations. Therefore, future spectroscopic observations of these objects are necessary for a better classification of these stars. 
On this matter, ongoing and future large scale spectroscopic surveys such as  The Large Sky Area Multi-Object Fiber Spectroscopic Telescope (LAMOST) \citep{2012RAA....12.1197C}, Sloan Digital Sky Survey V (SDSS-V) \citep{2017arXiv171103234K}, WEAVE \citep{2023MNRAS.tmp..715J} and 4MOST \citep{2019Msngr.175....3D} will be quite important to confirm the spectroscopic types of hot subdwarfs and to increase the size of the sample.
Furthermore, future photometric observations from space, such as TESS and the PLAnetary Transits and Oscillations mission \citep[PLATO,][]{2014ExA....38..249R,2018EPSC...12..969P}, or from ground-based initiatives such as Large Synoptic Survey Telescope (LSST) \citep{2019ApJ...873..111I},  BlackGEM \citep{2015ASPC..496..254B}, will contribute insights to the nature of the variabilities for the stars that are presented in this work. 

These surveys will allow us to create a volume-limited sample of sdBs. The initial efforts have been undertaken to create the first volume-limited sample of hot subdwarfs covering a range up to 500 pc (Dawson et al., submitted). 
Our forthcoming research will specifically concentrate on characterizing pulsating sdBs within 500 pc to determine the pulsation occurrence rate. 
Furthermore, these surveys will allow us to characterize the stars for which spectroscopic measurements are currently unavailable, as presented in Appendix Table \ref{known_tab}.
Every star featured in this study holds significant value as input for modeling purposes.
Including the findings presented here, the pulsational variability of 317 pulsating sdBs has been documented to date. However, only a small fraction of these stars has undergone in-depth asteroseismic analysis. 
Identifying common patterns in the pulsation spectrum in pulsating sdBs will guide our comparison of the measured frequencies to stellar models.
Therefore, developing a dedicated pipeline to do detailed seismic analysis for all pulsating sdBs presented in this work is crucial. 



\begin{acknowledgements}

M. U. gratefully acknowledges funding from the Research Foundation Flanders (FWO) by means of a junior postdoctoral fellowship (grant agreement No. 1247624N). 
I. P. acknowledges support a Royal Society University Research Fellowship (URF\textbackslash R1\textbackslash 231496).
H. D. is supported by the Deutsche Forschungsgemeinschaft (DFG) through grant GE2506/17-1.
P.N. acknowledges support from the Grant Agency of the Czech Republic (GA\v{C}R 22-34467S).
The Astronomical Institute in Ond\v{r}ejov is supported by the project RVO:67985815.
This research has used the services of \mbox{\url{www.Astroserver.org}} under reference W2MSWR.
The research has made use of TOPCAT, an interactive graphical viewer and editor for tabular data
Taylor \citep{TOPCAT_2005ASPC..347...29T}. This research made use of the SIMBAD database, operated at CDS, Strasbourg, France; the VizieR catalogue access tool, CDS, Strasbourg, France. 
This work has made use of data from the European Space Agency (ESA) mission Gaia (https://www.cosmos.esa.int/gaia), processed by the Gaia Data Processing and Analysis Consortium (DPAC, https://www.cosmos.esa.int/web/gaia/dpac/consortium). 
\end{acknowledgements}

\bibliographystyle{aa}
\bibliography{myrefs}

\appendix
\section{Tables and figures}
\label{appendix}
\subsection{Notes on objects in Table \ref{tab1} and \ref{tab2}}

Fourteen stars in \ref{tab1} were identified as intermediate (or GI) mode pulsators, with most, but three objects pulsating at intermediate frequencies higher than 160 d$^{-1}$. The three exceptions are:\\

 {\bf TIC 3905338}: recognized as a g-mode pulsator with consecutive frequencies extending from 47 up to 60 d$^{-1}$. Therefore, a few frequencies fall above the 55 d$^{-1}$ frequency limit considered in this paper as a common maximum frequency for g-mode pulsators (i.e., in agreement with all theoretical works). Interestingly, the light curve periodogram of this star shows a quite regular series of frequencies separated by 0.42 d$^{-1}$, which might be a result of frequency splittings of a few pulsation modes.

 {\bf TIC 241065253}: a g-mode pulsator also exhibiting two frequencies in the range between 85-95 d$^{-1}$. There is a clear $\sim$250-second period spacing visible between roughly seven g-modes, indicating $l_{1}$-mode excitation. Additionally, the light curve periodogram shows a signal at 0.714 d$^{-1}$, which might be assigned to an orbital frequency. However, this cannot be confirmed via visual inspection since the amplitude of the assumed orbital signal is low (most pulsating modes have higher amplitudes).

 {\bf TIC 325253096}: its periodogram shows just a single frequency at 76.8 d$^{-1}$.
\\
\\
Other interesting objects from Table \ref{tab1} are:
\\
 
 {\bf TIC 222892604}:  only two peaks are visible in the intermediate frequency range of the light curve periodogram. A frequency at 290.7124 d$^{-1}$ has an amplitude exceeding the detection threshold by more than 15 times, and it also shows an indication of splitting into a triplet.
 
 {\bf TIC 269766236}: the periodogram for this star shows clear period spacing of $\sim$250 seconds between 9 modes, indicating $l_{1}$-mode excitation.
 
 {\bf TIC 313007038}: identified as a GI mode pulsator. Its periodogram shows three g-mode signals of low amplitude between 9-25 d$^{-1}$ and three frequencies between 230 - 250 d$^{-1}$. A frequency at 248.36997 d$^{-1}$ has an amplitude exceeding almost 10 times the detection threshold.

{\bf TIC 441725813}: a star with rich and high-amplitude pulsation modes (compared to the detection threshold). The data for this star cover over 570 days across 14 seasons. There is an evident period spacing of approximately 260-270 seconds. The pulsation modes extend up to 390 d$^{-1}$ (visible in the ultra-short cadence data periodogram); therefore we identified the star as hybrid pulsator.
\\

At the end of Table \ref{tab1}, we present six p-mode pulsators found using ultra-short cadence data. Five of them exhibit a few (2 up to 4) pulsation frequencies in proximity to 600 d$^{-1}$; however, {\bf TIC 99499703} shows only a single frequency at 972.7577 d$^{-1}$.
\\

There is only a handful of objects on Table \ref{tab2} that might be more interesting than the others.\\

{\bf TIC 183799565}: the light curve periodogram for this star shows only a single frequency at 34.793 d$^{-1}$. If considered as the orbital frequency, it would correspond to one of the shortest orbital periods observed within sdB binaries. However, due to the low amplitude of this signal, a visual inspection of the phase-folded light curve reveals a large scatter, making it challenging to discern the shape of a possible binary light curve.
\\

{\bf TIC 165453703, TIC 239122172, TIC 241514378, TIC 443554222}: displaying an indication of asymmetrical phase-folded light curves and single frequencies in their periodograms, these stars are more likely to be pulsators than binaries.
\\

As mentioned in Section 3, stars with a low CROWDSAP factor (below 0.6) were examined to determine if they were responsible for the observed variability in the periodograms, as opposed to neighboring stars. However, in several cases, discerning the variability of a target or a neighboring star within the same pixel was challenging using only TPF file pixels. In such instances, we relied on the patterns of frequencies —referred to as an sdBV pulsation pattern, (primarily feasible only for objects from Table \ref{tab1})— visible in periodograms and the positions of neighboring stars from the TESS pipeline apertures in the HR diagram.  

We utilized HR diagrams based on Gaia data. In most cases, neighboring stars were situated on the main sequence in regions where sdBV pulsations cannot be observed, leading us to consider such variable targets as properly identified.

In the case of Table \ref{tab2}, out of 20 objects, four exhibit a low-CROWDSAP factor, namely {\bf TIC 95960421, TIC 239122172, TIC 270285517, and TIC 443554222}. Their periodograms reveal only single frequencies with low amplitudes. As a result, we were unable to confirm the variability of these stars, and further verification using ground-based and higher resolution telescopes is necessary.
\begin{table*}
\renewcommand{\arraystretch}{1}
\setlength{\tabcolsep}{11pt}
\centering
\caption{The list of 61 new sdBVs. Columns: 1) TIC number, 2),3) Right ascension, Declination, 4) TESS magnitude, 5) crowdness factor ($\tt{CROWDSAP}$), 6) number of TESS runs, 7) Pulsation type: G - gravity mode, I - intermediate mode, P - acoustic mode, H - hybrid mode, 8) notes. }
\label{tab1}
\begin{tabular}{rrrccrll}
\hline\hline
\noalign{\vskip 0.5mm}
~~~~~TIC & RA[deg] & DEC[deg] & $T_{\rm mag}$ & $\tt{CROWDSAP}$ & runs & Type & ~~~~Frequencies [d$^{-1}$], peaks~~~~ \\
\noalign{\vskip 0.2mm}
\hline
\noalign{\vskip 0.5mm}
    968226 & 328.82120 & -10.41736 & 13.91 &      0.99 &  1 &    I & 240-260, three peaks \\
   3905338 &  11.76393 & -11.87194 & 12.31 & 0.64-0.72 &  2 &   GI & 46-65, several \\
  11489563 &  18.04858 & -26.22435 & 13.33 & 0.98-0.99 &  2 &    G & 8-36, several \\
  14680532 & 174.37125 & -28.16202 & 12.81 & 0.84-0.89 &  3 &    G & 6-20, several \\
  16993518 & 235.24055 &  35.30954 & 15.41 &      0.93 &  1 &    I & 251-254, several \\
  17561485 & 117.04875 &  13.73037 & 15.72 & 0.31-0.39 &  4 &    G & 10-14, several \\
  27980141 & 279.21735 &  31.27523 & 14.79 & 0.39-0.54 &  3 &    G & 12-23, several \\
  29538272 & 287.75246 &  30.57915 & 14.40 & 0.42-0.44 &  2 &    G & 14-25, several\\
  35062562 & 180.44155 &  -3.76129 & 15.20 &      0.98 &  1 &    I & 160-200, several \\
  36707830 & 343.92995 &  -6.99441 & 12.46 &      0.99 &  1 &    G & 9-21, several \\
  43965472 & 343.99245 &  33.71994 & 12.85 &      0.74 &  1 &    G & 10-27, several \\
  44223583 & 345.63660 &  35.70738 & 14.35 &      0.65 &  1 &    G & 12-29, several \\
  60907492 & 139.96844 & -10.77548 & 15.33 & 0.60-0.61 &  2 &    G & 11-12, two peaks\\
  66493797 &  15.32322 & -33.71262 & 12.82 & 0.97-0.99 &  3 &    G & 13-20, several\\
  85145647 & 112.62083 &  62.19446 & 15.18 & 0.66-0.91 &  2 &    I & 195-255, several \\
 114196505 & 291.03594 & -12.97526 & 15.51 & 0.18-0.18 &  1 &    I & 228.8914, single (us) \\
 114423510 &  88.16722 &  19.82041 & 14.95 & 0.42-0.58 &  2 &    I & 280-290 single \\
 118032308 &   0.98168 &  27.81041 & 13.48 &      0.99 &  1 &    G & 11-23, several \\
 118297100 &  33.95948 &  29.34591 & 14.60 &      0.56 &  1 &    G & 17-25, three peaks \\
 138623536 &  96.77050 &  34.96937 & 14.44 & 0.59-0.72 &  4 &    G & 11-24, several \\
 152373379 &  65.51745 & -25.51679 & 13.83 & 0.84-0.92 &  3 &    G & 9-19, several \\
 153279970 &  92.90303 &   9.77157 & 14.30 &      0.44 &  1 &    G & 13-24, several \\
 159850392 & 269.16919 &  41.43028 & 14.43 & 0.65-0.79 &  3 &    G & 8-17, several \\
 170463675 & 122.36774 &  -8.17186 & 14.77 & 0.38-0.41 &  2 &    I & 190 single \\
 181914779 & 190.87520 & -14.73034 & 13.00 & 0.85-0.88 &  3 &    G & 19-23 two peaks \\
 188086093 & 161.59321 & -27.78143 & 15.03 & 0.94-0.96 &  2 &    G & 14-25 two peaks\\
 199715319 & 254.46497 &  55.19239 & 16.00 & 0.22-0.38 &  9 &    I & 242 single \\
 222892604 & 285.32903 &   6.24937 & 12.58 &      0.95 &  1 &    I & 282-292 two peaks\\
 229593795 & 273.03169 &  72.81189 & 15.44 & 0.34-0.71 & 28 &    G & 11-27 several \\
 229706981 & 278.00249 &  69.63906 & 15.80 & 0.26-0.66 & 26 &    G & 22-27 several \\
 233729964 & 277.44916 &  60.04192 & 15.86 & 0.73-0.92 & 28 &    G & 8-19 several \\
 241065253 &  21.76687 &  49.70961 & 11.35 & 0.89-0.94 &  3 &   GI & 7-30, 85-95 \\
 242176768 & 210.71375 & -47.08492 & 14.05 & 0.80-0.85 &  2 &    G & 10-29 several \\
 243366614 & 290.44122 &  57.39176 & 14.58 & 0.39-0.63 & 18 &    G & 9-26 several \\
 259126140 & 291.60482 &  69.88652 & 15.88 & 0.25-0.84 & 27 &    G & 18-30 several \\
 265787226 & 335.35344 &   2.27179 & 14.26 &      0.83 &  1 &    G & 11-24 several \\
 269766236 & 302.54188 & -36.20694 & 13.71 & 0.96-0.98 &  2 &    G & 15-32 several \\
 277773221 & 343.42604 & -66.90476 & 15.15 & 0.86-0.94 &  5 &    G & 13-24 several \\
 279464188 &  79.67630 &  70.99850 & 14.53 & 0.67-0.85 &  3 &    G & 8-26 several \\
 292467033 &  32.38524 &  43.12014 & 14.49 &      0.71 &  1 &    G & 22-25 two peaks\\
 313007038 & 137.48105 &  73.31281 & 15.20 & 0.82-0.96 &  5 &    G & 17-20 two peaks \\
 325253096 & 176.64679 &   0.20932 & 15.64 &      0.88 &  1 &    I & 76.80 single \\
 331656308 &  70.16249 &  76.05731 & 14.68 & 0.55-0.68 &  3 &   GI & 9-25, 230-250 several \\
 332841294 & 180.60843 & -51.91663 & 13.77 & 0.50-0.53 &  2 &    G & 13-25 several \\
 340223812 & 201.01250 & -71.05591 & 14.20 & 0.45-0.51 &  2 &    G & 11-33 several \\
 381203990 &  15.30557 &  31.43208 & 13.20 &      0.98 &  1 &    G & 13-30 several \\
 392092589 & 120.36491 &  25.96193 & 15.02 & 0.40-0.74 &  3 &   GI & 18-28, 230 several \\
 392722474 & 288.98207 &  32.35205 & 14.27 & 0.34-0.36 &  2 &    G & 7-21 several \\
 396615028 &  39.98837 &  42.96686 & 14.00 &      0.81 &  1 &    G & 14-30 several \\
 399171956 & 302.46528 &   3.17550 & 12.60 &      0.96 &  1 &    G & 16-30 several \\
 429807453 &  94.48614 &  18.83019 & 13.82 & 0.56-0.61 &  3 &    G & 14-28 several \\
 440435901 &  90.21970 &  47.16901 & 14.44 &      0.83 &  1 &    G & 11-13 two \\
 441725813 & 256.24324 &  73.07867 & 11.05 & 0.99-1.00 & 14 &    H & 3-45 +peaks up to 400 (us) \\
 457225725 &  75.86300 &   2.12333 & 14.59 &      0.96 &  1 &    G & 10-21 two \\
 461658287 & 124.03295 &  48.06381 & 15.28 & 0.79-0.89 &  3 &    G & 12-15 two \\
\hline
\noalign{\vskip 0.2mm}
   4632676 & 180.92172 &  25.51987 & 15.04 & 0.91-0.91 &  1 &    P & 628.9127 single (us) \\
  63168679 & 118.37131 &  23.41008 & 13.84 & 0.80-0.84 &  2 &    P & 580-650 several (us) \\
  99499703 & 317.83761 & -23.80389 & 14.61 & 0.46-0.46 &  1 &    P & 972.7577 single(us) \\
 178081355 &  49.59612 &  41.92271 & 14.58 & 0.32-0.32 &  1 &    P & 500-600, two peaks (us) \\
 357232133 & 115.45486 &  55.41408 & 14.56 & 0.92-0.95 &  2 &    P & 590-700, three peaks (us) \\
 364966239 & 298.21182 &  58.14364 & 14.88 & 0.55-0.71 &  7 &    P & 560-600, two peaks (us) \\
 
\hline
\end{tabular}
\end{table*}

\begin{figure*}
\centering
        \includegraphics[width=0.95\linewidth]
        {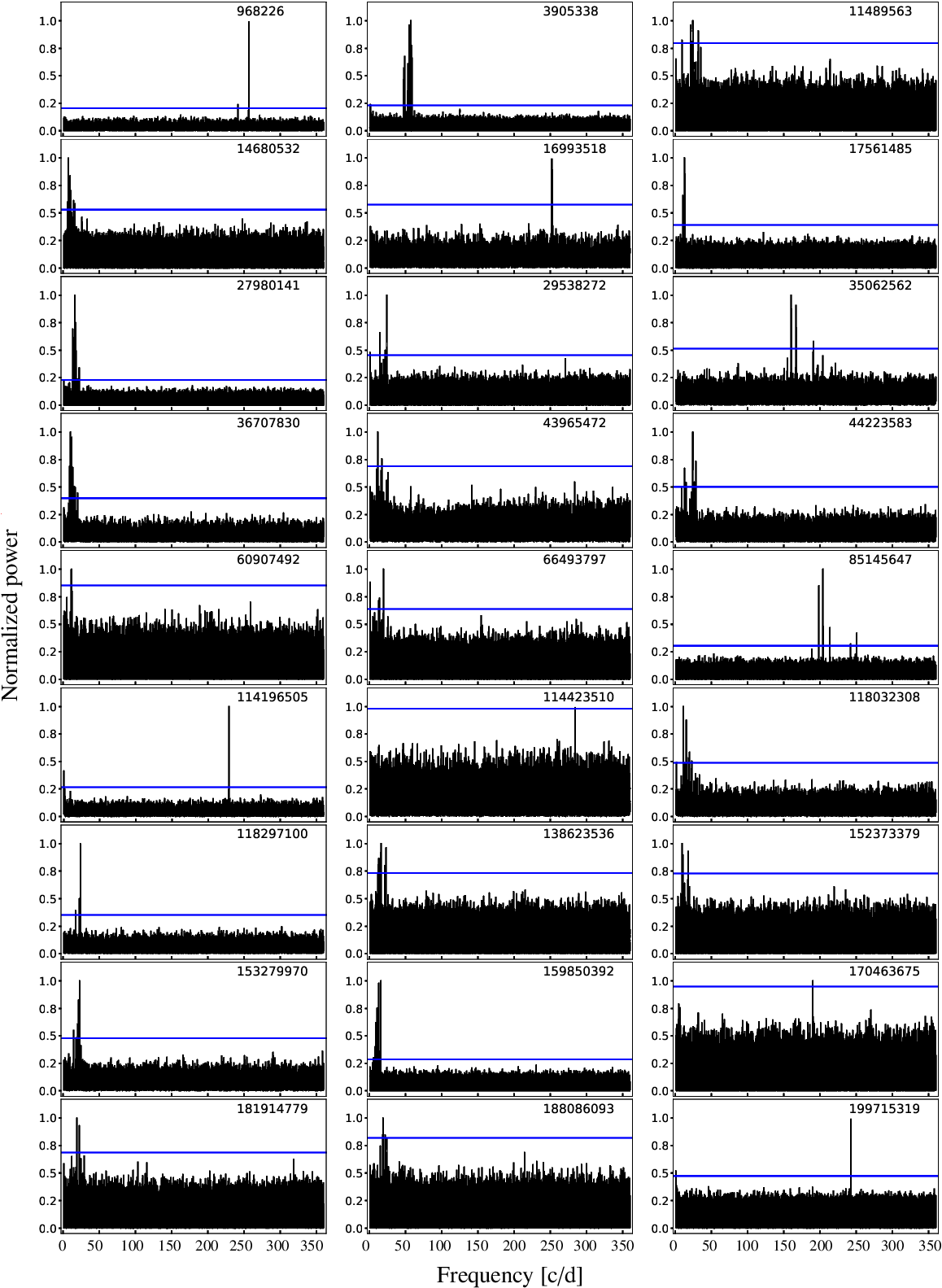}
\hfill

\caption{Lomb-Scargle periodograms of 61 newly found sdBV stars (27 objects from table\,\ref{tab1} are shown). Blue horizontal lines are 5\,$\sigma$ detection thresholds.} 

\label{fig:Periodograms_a}
\end{figure*}

\begin{figure*}
\centering
        \includegraphics[width=0.95\linewidth]
        {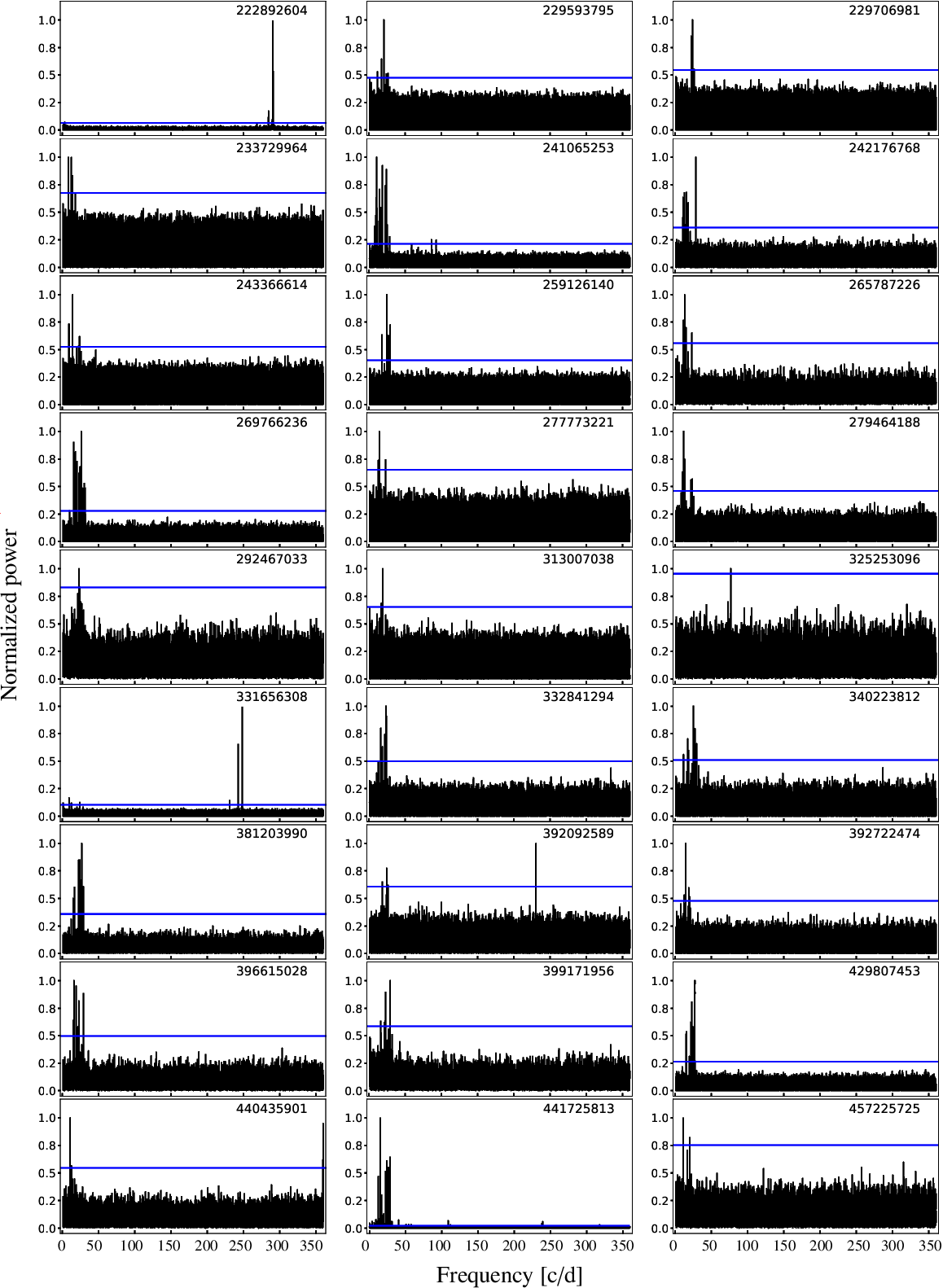}
\hfill

\caption{Continuation of Fig. \ref{fig:Periodograms_a}: Lomb-Scargle periodograms of the next 27 sdBV stars from table\,\ref{tab1}. Blue horizontal lines are 5\,$\sigma$ detection thresholds.} 

\label{fig:Periodograms_b}
\end{figure*}

\begin{figure*}
\centering
        \includegraphics[width=0.95\linewidth]
        {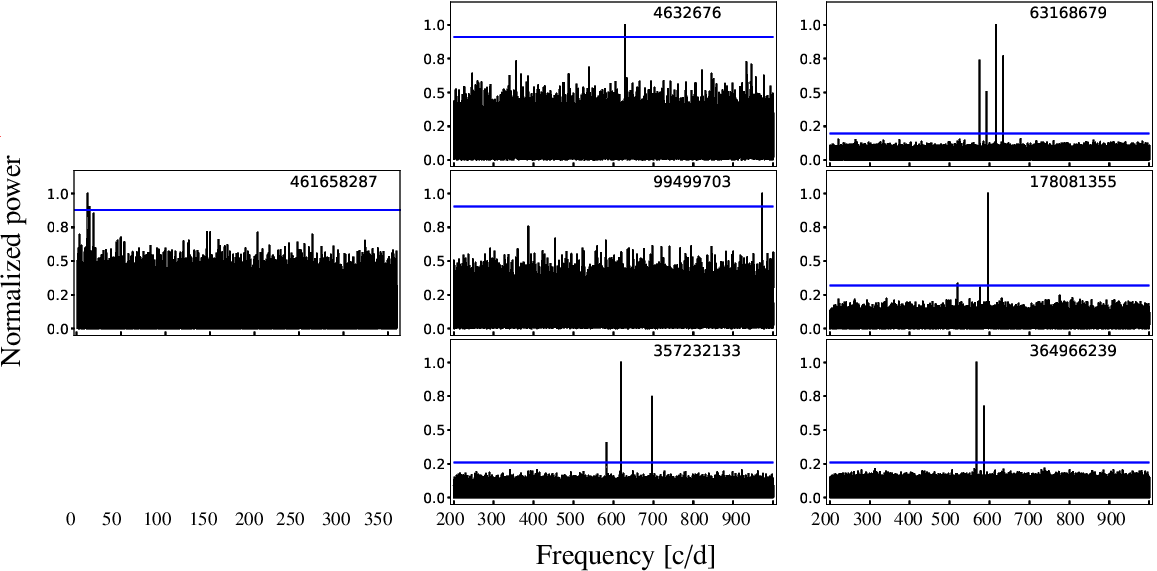}
\hfill

\caption{Continuation of Fig. \ref{fig:Periodograms_a}: Lomb-Scargle periodograms of the last sdBV (lef side) from the top part of table\,\ref{tab1} and 6 p-mode pulsators found in the ultrashort cadence (20 sec integration time) data. Blue horizontal lines are 5\,$\sigma$ detection thresholds.
\\
\\
\\
} 

\label{fig:Periodograms_p}
\end{figure*}

\begin{table*}
\renewcommand{\arraystretch}{1}
\setlength{\tabcolsep}{10pt}
\centering
\caption{The list of 20 sdB stars of uncertain variability type. Columns: 1) TIC number, 
2),3) Right ascension and Declination, 4) TESS magnitude, 5) CROWDNESS factor ($\tt{CROWDSAP}$), 6) number of TESS runs, 7) variability type: B - binary, R - rotation, G - gravity mode, 8) notes.}
\label{tab2}
\begin{tabular}{rrrccrll}
\hline\hline
\noalign{\vskip 0.5mm}
~~~~~TIC & RA[deg] & DEC[deg] & $T_{\rm mag}$ & $\tt{CROWDSAP}$ & runs & Type & Frequencies in [d$^{-1}$]) \\
\noalign{\vskip 0.5mm}
\hline
\noalign{\vskip 2mm}
4037      & 218.95000 & -24.90300 & 15.57 & 0.63      &  1 & B?G?   & 5.53 single, no sub- or harm.  \\
14346839  & 124.30132 &  22.91486 & 14.55 & 0.83-0.96 &  3 & B?G?   & 23.37 single, no sub- or harm.  \\
58873368  &  33.79626 &  15.00116 & 14.59 & 0.99-0.99 &  2 & B?G? & 10.08, single, no sub- or harm. \\
95960421  & 141.30886 &  -9.58161 & 15.71 & 0.28-0.33 &  2 & B?G?   & 29.72 single, no sub- or harm.  \\
125556577 & 114.92392 & -27.45612 & 13.79 & 0.63-0.64 &  3 & B?G?   & 6.06 single, no sub- or harm.  \\
165453703 & 201.00302 &  49.37564 & 12.72 & 0.97-0.98 &  3 & G?B? & 7.23 single, no sub- or harm.\\
167554511 & 101.20387 & -60.41036 & 14.67 & 0.83-0.92 & 30 & B?G?   & 6.63 single, no sub- or harm.  \\
\end{tabular}
\end{table*}

\begin{figure*}
\centering
        \includegraphics[width=0.95\linewidth]
        {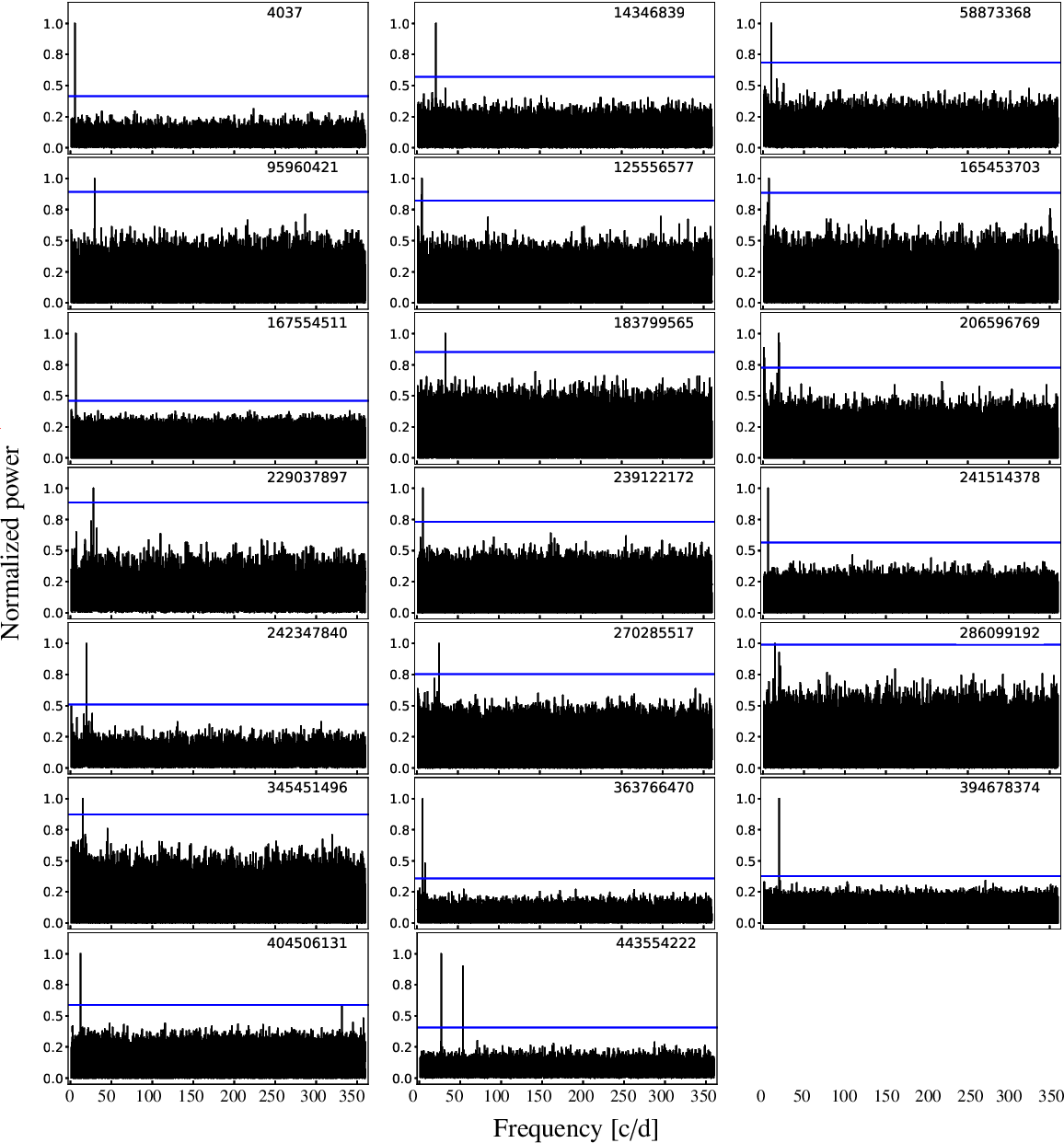}
\hfill

\caption{Periodograms of 20 sdB stars from table\,{\ref{tab2}} of uncertain variability. Blue horizontal lines are 5\,$\sigma$ detection thresholds.} 
\label{fig:Periodograms_c}
\end{figure*}

\onecolumn

\begin{landscape}

\small\tabcolsep=2.5pt

\centering

\begin{longtable}{ccccccccc}

\caption{Catalogue of all 256 known pulsating hot subdwarfs.} \\

\toprule
                    TIC &  Name &  RA[deg] &  DEC[deg] &  $T_{\rm eff}$ (K)  & $\log{g}$ (cgs)  & $\log{(n_{\rm He} / n_{\rm H})}$ & Distance (pc)   & Reference                \\                           
\midrule
\endfirsthead

\toprule
                       TIC &  Name &  RA[deg] &  DEC[deg] &  $T_{\rm eff}$ (K) & $\log{g}$ (cgs) & $\log{(n_{\rm He} / n_{\rm H})}$ & Distance (pc)   & Reference              \\  
\midrule
\endhead
\midrule
\multicolumn{9}{r}{{Continued on next page}} \\
\midrule
\endfoot

\bottomrule
\endlastfoot

 138707823 &                         Ton S 135 &   0.8420 & -23.6495 &   25000 $\pm$ 1250.0 &   5.6 $\pm$  0.2 &  -2.3 $\pm$  0.2 &  -       &                     1986A\&A...155...33H \\
 201251043 &                      CPD-55 10148 &   0.8889 & -54.7169 &                    - &     - &     - &  $ 387^{  +5}_{   -6}$      &                     2020yCat.1350....0G \\
 322009509 &                            JL 166 &   3.1793 & -45.8543 &   34400 $\pm$      - &  5.75 $\pm$    - &  -0.8 $\pm$    - &  $2307^{   +6}_{  -7}$      &                     2010A\&A...513A...6O \\
 437746793 &                       PG 0011+283 &   3.5927 &  28.6155 &   24938 $\pm$  298.0 &  5.35 $\pm$ 0.04 & -3.77 $\pm$ 0.15 &  $ 374^{   +5}_{  -9}$      &                     2021ApJS..256...28L \\
 405212419 &                       PG 0014+068 &   4.2261 &   7.0750 &   34100 $\pm$      - &  5.77 $\pm$    - & -1.68 $\pm$    - &  $1811^{   +6}_{  -7}$      &                     2010A\&A...513A...6O \\
  80290366 &                       CD-48   106 &   7.9235 & -47.4222 &   25200 $\pm$ 1200.0 &   5.2 $\pm$  0.2 &  -1.8 $\pm$    - &  $ 467^{   +5}_{  -7}$      &                     1984A\&A...130..119H \\
 191442416 &                      HS 0039+4302 &  10.6767 &  43.3169 &   32300 $\pm$  800.0 &   5.7 $\pm$  0.1 &  -2.3 $\pm$  0.2 &  $1326^{   +6}_{  -8}$      &                     2003A\&A...400..939E \\
  80427831 &                     EC 00404-4429 &  10.7016 & -44.2237 &                    - &     - &     - &  $ 641^{ -15}_{  +19}$      &                     2020yCat.1350....0G \\
\end{longtable}

\end{landscape}

\end{document}